\documentclass[fleqn,usenatbib]{mnras}

\usepackage{newtxtext,newtxmath}

\usepackage[T1]{fontenc}
\usepackage[latin1]{inputenc}
\usepackage{ae,aecompl}

\usepackage{graphicx}	
\usepackage{amsmath}	
\usepackage{amssymb}	

\title[Constraining dark photon]{Constraining dark photon properties with Asteroseismology}

\author[A. Ayala et al.]{
Adri\'an Ayala$^{1}$\thanks{E-mail:aayala@ugr.es)},
Ilidio Lopes$^{2,3},$
Antonio Garc\'{\i}a Hern\'andez$^{4,1},$
Juan Carlos Su\'arez$^{4,1},$
\newauthor
\'{I}\~nigo Mu\~noz Elorza$^{5},$
\\
$^{1}$Instituto de Astrof\'{\i}sica de Andaluc\'{\i}a (CSIC). Glorieta de la Astronom\'{\i}a s/n. 18008, Granada, Spain\\
$^{2}$Centro de Astrof\'{\i}sica e Gravita\c c\~ao  - CENTRA, 
	Departamento de F\'{\i}sica, Instituto Superior T\'ecnico - IST,
	Universidade de Lisboa - UL,\\ Av. Rovisco Pais 1, 1049-001 Lisboa \\
$^{3}$Institut d'Astrophysique de Paris, UMR 7095 CNRS, Universit\'e Pierre et Marie Curie, 98 bis Boulevard Arago, Paris 75014, France\\
$^{4}$Departamento de F\'{\i}sica Te\'orica y del Cosmos, Universidad de Granada, Campus de Fuentenueva s/n, 18071, Granada, Spain\\
$^{5}$ European Astronaut Centre/ESA, Cologne (Germany)\\
}

\date{Accepted XXX. Received YYY; in original form ZZZ}

\pubyear{2019}

\begin{document}
\label{firstpage}
\pagerange{\pageref{firstpage}--\pageref{lastpage}}
\maketitle

\begin{abstract}
 Dark photons are particles invoked in some extensions of the Standard Model which could  account for at least part of the dark matter content of the Universe. It has been proposed that the production of dark photons in stellar interiors could happen at a rate that depends on both, the dark photon mass and its coupling to Standard Model particles (the kinetic mixing parameter $\chi$).
 In this work we aim at exploring the impact of dark photon productions in the stellar core of solar mass RGB stars during late evolutionary phases. We demonstrate that near the so-called RGB bump, dark photons production may be an energy sink for the star sufficiently significative to modify the extension of the star convective zones. We show that Asteroseismology is able to detect such variations in the structure, allowing us to predict an upper limit of $\rm 900\ eV$ and $5\times 10^{-15}$ for the mass and kinetic mixing of the dark photons, respectively. We also demonstrate that additional constraints can be derived from the fact that dark photons increase the luminosity of the RGB tip over the current observational uncertainties. This work thus paves the way for an empirical approach to deepen the study of such dark-matter particles.

 \end{abstract}



\section{Introduction}
 
 Despite its success, the Standard Model of Particle Physics (SM) leaves many questions unanswered, suggesting that it should be extended. Dark matter, dark energy, and the CP (charge-parity) violation are some of these open questions. Many of the solutions proposed to solve the dark matter problem consist of extensions to  the SM by assuming the existence of a dark sector, made of new particles which may or may not interact with the known fundamental particles. These hypothetical particles have been recognized as affecting astrophysical observables \citep{1999AIPC..490..125R}. Therefore, some constraints have been derived, from astrophysical observations, on Weakly Interacting Massive Particles or WIMPS \citep{2011PhRvD..83f3521L,2011ApJ...729...51S,2009MNRAS.394...82S,2008PhRvL.100e1101S}, as well as on the so-called Weakly Interacting Slim Particles (WISPs) such as axions \citep{2010ARNPS..60..405J,1978PhRvL..40..223W, 1978PhRvL..40..279W}. Among the effects of these particles there are changes in the standard stellar evolution.\par 
  For instance, it was pointed out some years ago that the post-main sequence stars could be affected by axion-photon interactions mediated by means of the Primakoff process in stellar interiors. This changes the lifetime of the central helium burning, during the horizontal branch (HB) phase  \citep{2012JCAP...02..033H, 1987PhRvD..36.2211R}. More specifically, Primakoff axion emission from HB stars modifies the $R$ parameter, an observable relying on HB lifetimes and related to globular clusters, which has been used to constrain axion-photon coupling \citep{2014PhRvL.113s1302A}.\par
The so-called tip of the red giant branch (TRGB), a local luminosity maximum in HR diagram which takes place by the end of RGB phase, has been used to derive bounds on axions and neutrino properties \citep{2013PhRvL.111w1301V}. Also occurring during the RGB phase, the RGB bump, a maximum of the luminosity function that happens when the hydrogen shell reaches the chemical gradient left by the convective envelope \citep{1985ApJ...299..674K, 1968Natur.220..143I, 1967ZA.....67..420T} has motivated a lot of research. After its first detection \citep{1985ApJ...299..674K}, a number of studies mentioned the discrepancies between theoretical models and the bump luminosity, arguing for the necessity of some additional mechanism. Specially that concerning convection or the additional cooling by new particles \citep{2018ApJ...859..156K,2006ApJ...641.1102B, 2003A&A...410..553R, 1999ApJ...518L..49Z, 1997MNRAS.285..593C, 1990A&A...238...95F}.\par
 In this work we suggest the study of the RGB bump in order to constrain the properties of beyond Standadard Model particles. To do that we introduce, in the simulation of the RGB stars, an additional energy loss, the loss consisting of the emission from stellar cores of dark photons, boson particles which have been invoked in some minimal modifications of SM and could be a candidate for at least part of the dark matter content in the Universe \citep{2019PhRvD..99l3511K,2019PhRvL.122t1801B,2014PhRvD..90c5022F, Redondo:2013lna,2012JCAP...06..013A, 2009JCAP...02..005R,1986paun.conf..133H, 1982ZhETF..83..892O}.
 We explored the changes in the internal structure of RGB stars due to the dark photon emission from their degenerate cores. Then we performed a study of the observational consequences from the point of view of Asteroseismology, the study of oscillations in star light curves originated by internal instabilities (see e.g.,  \citet{2010aste.book.....A}). This technique is nowadays the most accurate way of looking into the stellar interiors. In order to compare the prediction of asteroseismic parameters of the dark photon models with the observations we used the characterization of the RGB bump stars previously done by \citet{2018ApJ...859..156K}, a reference work for this paper, with data coming from the APOKASC catalogue \citep{0067-0049-233-2-25}.\par
 In addition, due to the fact the dark photon cooling can affect the He core mass, as we demonstrate in
 this paper, it is appealing to study the luminosity at
 the tip of the RGB. This observable depends on the He mass by the end of the RGB, and therefore could be an additional constraint on dark photon models.\par

 This paper is organized as follows: in section 2 we introduce the theoretical motivations of dark photons and the expression of the emission rate, as well as the possible effects of dark photons on the stellar oscillations. Section 3 deals with the results of the simulations concerning the stellar structure of the equilibrium models. The changes on the pulsations of the stars are described in Section 4. Section 5 deals with the RGB tip luminosity as an additional constraint on dark photons. Finally, section 6 summarizes the conclusions and addresses future work.\par


%


\section{Physical case for dark photon}
Dark photon was proposed as the associated particle of a vector field, $B$, which can couple to
 the hypercharge of Standard Model, by means of the so-called kinetic mixing \citep{1986paun.conf..133H, 1982ZhETF..83..892O}. The SM lagrangian  is then modify \citep{Redondo:2013lna}
 as follows:
\begin{equation}
 \mathcal{L} = -\frac{1}{4} A_{\mu\nu}A^{\mu\nu} -\frac{1}{4} B_{\mu\nu}B^{\mu\nu} - \frac{\chi}{2}A_{\mu\nu}B^{\mu\nu}+\frac{m_{V}^2}{2}B_{\mu}B^{\mu}
\end{equation}

where $m_{V}$ and $\chi$ stand, respectively, for the mass of the dark photon and the kinetic mixing or coupling between the photon, $A$, and $B$ fields. \par

Several experiments are devoted to dark photon search \citep {2018PhRvD..98c5006B, 2018JHEP...07..094B,2018arXiv180410777D,2018arXiv180708516X,2015JCAP...09..042S, 2014arXiv1410.0200D}. More specifically
some recent experimental constraints on dark photon parameters are those of \citet{2019JCAP...07..008K} and \citet{2018PhLB..787..153A}. In 
addition, constraints on dark photon models
have been put on the basis of astrophysical arguments \citep{2019JCAP...05..015G,2019arXiv190205962G,2018arXiv180901139K, 2018JCAP...03..043C, 2014arXiv1412.8378A}. The fact that under certain conditions of the stellar interiors, 
dark photons could be emitted, carrying energy, with a rate proportional to $\chi$ and $m_V$ has been exploited by some authors to derive bounds on both parameters from the main sequence or the post-main sequence star features
\citep{2015JCAP...10..015V, 2013PhLB..725..190A, Redondo:2013lna,2010arXiv1010.4689C}. With the exception of \cite{2015JCAP...10..015V}, these previous works do not take into account the feedback of 
dark photon emissions on the stellar evolution, as far as they depart from a previous stellar model and calculate an average of the energy loss due to dark photons. Then 
the possible effects on stellar structure due to dark photon emissions are neglected .\par

According to previous research the flux of hypothetical dark photons from stellar interiors has been
computed under the assumption that these particles interact only with normal matter through kinetic mixing, by means of longitudinal and transverse modes of plasmon oscillations \citep{Redondo:2013lna,2008JCAP...07..008R}.
For longitudinal plasmon oscillation the fact that the dispersion relation equals the plasma frequency ensures the existence of a region inside the stars where the kinetic mixing between the photons and dark photons reaches a maximum, leading to a resonance \citep{Redondo:2013lna}. On the basis of this result, we restrict the study to longitudinal (L) modes. Now in case of longitudinal mode dark photon
emission, according to \citet{2013PhLB..725..190A}, the rate can be expressed in the form 
\begin{equation}
\epsilon_{L}= \frac{\chi^2m_{V}^2}{2\pi^2\rho} \int^{\infty}_{m_{V}} \frac{\omega^4\sqrt{\omega^2-m_{V}^2}}{e^{\frac{\omega}{T}}-1}
\frac{\Gamma_{L}}{\left(\omega^2-\omega_{P}^2\right)^2 +\left(\omega\Gamma_{L}\right)^2}\  d\omega\, .
\label{eq:rate}
\end{equation}
 Where as usual $\rho$ and $T$ stand for density and temperature, $\Gamma_{L}$ for the rate of longitudinal plasmon oscillation, $\omega$ for the frequency associates to dark photon dispersion relation, $\omega_{P}$ is the plasma frequency, $m_{V}$ the dark photon mass, and $\chi$ the kinetic mixing, which is a non-dimensional parameter. Then, if we consider resonant emission of dark photon, by taking the limit
\begin{equation}
\lim\limits_{\omega\Gamma_{L}\rightarrow 0} \frac{\omega \Gamma_{L}}{\left(\omega^2-\omega_{P}^2\right)^2 +\left(\omega\Gamma_{L}\right)^2} \approx \frac{\pi}{2\omega_{P}}\delta\left(\omega-\omega_{P}\right) , 
\label{eq:limit}
\end{equation}
and plug this into Equation ~\ref{eq:rate}, after including the change of units in numerical constants, we obtain the resonant expression of the rate, depending on $\chi$ and $m_{V}$.

\begin{equation}
\epsilon_{L} = \frac{1.62 \times 10^4 \chi^2 m_{eV}^2 x_{p}^2 T_{K}^3 \sqrt{x_{p}^2-x_{m}^2}}{\rho_{cg}\left(e^{\frac{\omega_{P}}{T}}-1\right)} \left(\frac{erg}{g \cdot s}\right)
\label{eq:rater}
\end{equation}
 
Where $m_{eV}$ is $m_{V}$ in eV units, $x_{p}=\omega_{P}/T$ and $x_{m}=m_{V}/T$.\\
 
For dark photon masses up to $200$ keV, the condition $x_{p}-x_{m} > 0$ holds for the typical values of $\omega_{P}$ and T of RGB stars with initial masses around the solar one \citep{2014arXiv1412.8378A}, 
thus the expression under the square root is positive and the rate is real. This condition motivated us to look into the dark photons throughout the RGB phase. As in the dark photon mass region $\rm \sim 1\, keV$, 
constraints from RGB stars could be more reliable
than those coming from dark matter experiments,
finally we decided to explore the mass range between $\rm 300\ eV$ and $\rm 900\ eV$, as well as a kinetic mixing in the range of $2 \times 10^{-15}-9 \times 10^{-15}$, not excluded by previous research in this mass interval.\par

\subsection{Asteroseismology as dark photons probe} 
 
 Asteroseismology is living a golden era thanks to space missions devoted to ultra-precise photometry, such as {\it{CoRoT}} \citep{2006ESASP.624E..34B}, {\it{Kepler}} \citep{2010ApJ...713L..79K} and, recently, {\it{TESS}} \citep{2015JATIS...1a4003R}. Stellar oscillations strongly depend on the stellar structure so Asteroseismology is able to probe interiors of pulsating stars. Bulk stellar parameters are inferred with an uncertainty of up to 2\% on the radius, 4\% on the mass and 15\% on the age for stars showing solar-like oscillations \citep{2017ApJ...835..173S}. The oscillation modes on these stars form a clear structure where modes of consecutive radial orders and the same spherical degree are separated by an almost constant frequency distance. This periodic spacing is known as the large separation, $\Delta\nu$ \citep{2010aste.book.....A}, and it is related to the mean stellar density.
\\

 In addition, Asteroseismology also succeeds in constraining the phenomenology of new particles in
stars, as these particles act as supplementary sources or sinks of energy, or they imply a novel transport mechanism inside stars \citep{2013ApJ...765L..21C, 1994A&A...290..845L}. In that case the internal physical condition and the gradients vary. Therefore, the internal structure, namely the extension of radiative or convective zones, is modified and with them the sizes of the internal cavities the oscillations propagate throughout. According to this idea, the dark photon emission is able to change the internal structure of the star during its evolution, having an impact on the oscillation frequencies of RGB stars. Thus, the analysis of the periodic spacings in the frequency content of RGB stars has the potential to impose constraints on the characteristics of dark photons. This is the aim of our analysis: compare equilibrium and oscillation models with observations in order to constrain dark  photon emission.\par

RGB stars excite solar-like oscillations.Throughout the RGB phase the degenerate core is dominated by gravity modes (g modes), whereas, in the outer regions, pressure modes (p modes) have a much larger amplitude \citep{2017MNRAS.466.3344E,2011Natur.471..608B}. Pressure modes carry the information of $\Delta\nu$. The determination of the frequencies for modes of a given radial and angular order, $\nu_{ln}$, makes it possible to study second differences, $\delta_{2}\nu_{ln}$ \citep{1994A&A...290..845L, 1990LNP...367..283G}, defined as

\begin{equation}
 \delta_{2}\nu_{ln} = \nu_{l,n+1} - 2\nu_{l,n} + \nu_{l,n-1} ,
 \label{eq:delta02}
\end{equation}

which tests the convective boundaries in stars of masses $\rm 1\ M_{\odot} < M < 1.5\ M_{\odot}$. In this work we deal with the variation of both parameters, second difference for radial modes ($\delta_{2}\nu_{0n}$), and $\Delta\nu$, produced by the inclusion of dark photons.\par


\section{The equilibrium models}
 In order to compute the stellar models we used MESA, an open-source 1D code which includes most of the relevant physics, like updated opacities and convection \citep{2019ApJS..243...10P, 2018ApJS..234...34P, 2015ApJS..220...15P, 2013ApJS..208....4P, 2011ApJS..192....3P}. We extended the code to take into account the additional energy sink of dark photons, previously estimated by \citet{2013PhLB..725..190A} and \citet{Redondo:2013lna}, implementing Equation~\ref{eq:rater}.\par

We computed a grid of stellar models with $\rm 1.0\ M_{\odot}$, $\rm 1.2\ M_{\odot}$, $\rm 1.4\ M_{\odot}$, and solar metallicity, setting the mixing length $\alpha = 1.9658$. The initial stellar masses and metallicity were set to these values to compare the models with the observations, as we do in the last part of this work. \citet{2018ApJ...859..156K} have looked into the effect of overshooting from the base of the convective envelope on the bump. Notwithstanding  overshooting is poorly constrained by theory, within the exponentially decaying overshooting formalism used in MESA, they exclude values higher than $\alpha_{ov}=0.05$, on the basis of the theoretical prediction for the RGB bump average large separation. In fact, for overshooting values higher than $\alpha_{ov}=0.05$ there is an offset between the theoretical values and the observations, which is pretty much over the $2\sigma$ limits. On the contrary, the authors find that a moderate overshooting, $\alpha_{ov} = 0.025$, can address the mismatch between observations and the RGB bump models. We have introduced this last $\alpha_{ov}$ value in our simulations with the aims of checking if overshooting and dark photon effects can be disentangled.\par

 To study the effects of the dark photon in the models, we computed evolutionary tracks, varying the values of $\chi$ between $2\times 10^{-15}$ and $9\times 10^{-15}$ and the dark photon mass, $m_{V}$ between 300 and 900 $\rm eV$. We chose these values because they are not excluded by previous astrophysical research, even though the lower masses are close to the bounds imposed by the solar lifetime \citep{2015JCAP...09..042S}. Moreover, the resonance condition discussed before is satisfied for all the mass values.\par

For each model, we located the RGB bump searching the luminosity interval along the RGB that is crossed three times, as a consequence of the passage of the hydrogen-burning shell through the discontinuity in chemical composition left over by the first dredge-up \citep{2016A&A...585A.124C}. A local maximum and minimum in the evolutionary track is characteristic of the RGB bump. We have focused on the minimum in this paper, referring all the calculations to that point.\par
 
\begin{figure}
    \centering
    \includegraphics[width=0.50\textwidth]{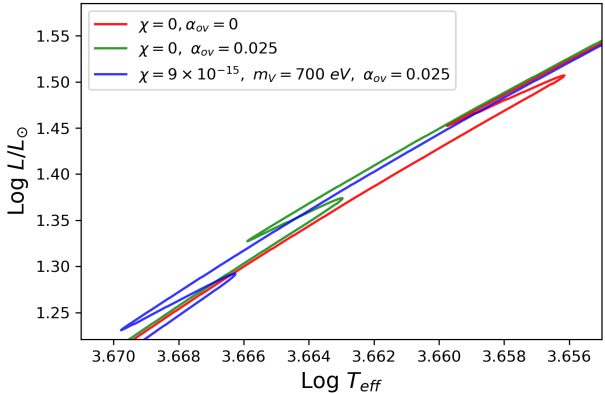}
    \caption[RGB bump position in HR diagram]{RGB bump position in the HR diagrams, for models with neither dark photons nor overshooting (red), overshooting $\alpha = 0.025$ (green), and $\rm m_{V}=700\ eV$ and $\chi=9 \times 10^{-15}$ with  $\alpha = 0.025$ (blue).}
    \label{fig:HR}
\end{figure}

 The ages, effective temperatures and luminosities at bump for a reference model (without neither dark photons nor overshooting), the models regarding dark photons, overshooting $\alpha_{ov,env} = 0.025$, and the combined effects of dark photons and overshooting are shown in Table~\ref{tab:table1}. It is noticeable that the bump occurs earlier with respect to the standard scenario when dark photons production is considered. On the other hand, a slight increase in effective temperature, proportional to $\chi$ and $m_{V}$ is also verified. In addition, we report a decrease of luminosity as $\chi$ and $m_V$ are enhanced. Therefore, the introduction of dark photons implies a displacement of the position of the bump in the HR diagram, as it is shown in Figure~\ref{fig:HR}. The overshooting effects on the age, luminosity and effective temperature at bump are analogous to those of dark photons: the bump is reached earlier, the effective temperature increases whereas the luminosity decreases. When overshooting and dark photon occur simultaneously, the variations with respect to the reference model are larger, reaching a maximum for the model $\alpha_{ov,env} = 0.025$, $m_V = 900
 \ eV$ and $\chi = 9\times 10^{-15}$.\par

The effects of dark photon cooling and overshooting on the He core and the hydrogen shell, occurring throughout the RGB phase, are shown in Table~\ref{tab:table2}. The mass of the helium core decreases with respect to the model with $\chi = 0$, as does the central temperature. We also recall that the nuclear energy rate due to the pp and CNO reactions within the hydrogen shell is reduced when the dark photon production increases. These effects are compatible with dark photon playing the role of an additional energy sink throughout the RGB. It is remarkable that, also in this case, the envelope overshooting reinforces the dark photon effects, and that the highest variations of helium mass core and central temperature correspond to the model $\rm m_V = 900\ eV$ and $\rm \chi = 9\times 10^{-15}$ whith overshooting.


\begin{table}
  \begin{center}
    \caption{Values of Age, Log $\rm T_{eff}$ and Log $\rm L/L_{\odot}$ for models with $1.0\ M_{\odot}$, $\rm Z_{\odot}$, different dark photon parameters and $\alpha_{ov} = 0, 0.025$ at the RGB bump.}
    \label{tab:table1}
    \begin{tabular}{c|c|c|c|c|c} 
     \textbf{$\alpha_{ov}$} & \textbf{$\chi$} & $\rm m_{V}\ (eV)$ & Age (Gyr) & \textbf{$\rm Log\ T_{eff}$} &\textbf{$\rm Log\ L/L_{\odot}$}\\
      
      \hline
      
      $0$ & $0$ & 0 & 12.25 & 3.660 & 1.452 \\
      
      $0$ & $2\times 10^{-15}$ &900& 12.23 &3.660 &1.441 \\
      
      $0$ & $7\times 10^{-15}$ &900 & 12.08& 3.664& 1.365 \\
      
      $0$ & $9\times 10^{-15}$ &900 &12.02  &3.665& 1.338\\
      $0$ & $5\times 10^{-15}$ &500 &12.21   &3.661 & 1.431 \\
      $0$ & $5\times 10^{-15}$ &700 & 12.18  & 3.662& 1.414 \\
     $0$ &  $5\times 10^{-15}$ &900 & 12.15  & 3.662  & 1.397\\
     $0.025$ & $0$ & $0$ &12.24  &3.666 & 1.327\\
     $0.025$ & $2\times 10^{-15}$ &900& 12.22 &3.666 &1.315 \\
      
      $0.025$ & $7\times 10^{-15}$ &900 &12.07  & 3.669& 1.232 \\
      
      $0.025$ & $9\times 10^{-15}$ &900 &12.01  &3.671 & 1.204\\
      $0.025$ & $5\times 10^{-15}$ &500 &12.19   &3.667 & 1.304 \\
      $0.025$ & $5\times 10^{-15}$ &700 &  12.17 & 3.668& 1.286\\
     $0.025$ &  $5\times 10^{-15}$ &900 &  12.14  & 3.668  & 1.267\\
     \end{tabular}
  \end{center}
\end{table}

The reduction of luminosity shown in Table~\ref{tab:table1} is in agreement with the contraction of the whole star, indicated by the ratio between the mean and the central densities, $\rho/\rho_{c}$ in Table~\ref{tab:table3}. Nevertheless, most relevant for the internal structure is the deeper penetration of the convective zone (see also Table~\ref{tab:table3}), which reaches lower regions than in the standard evolutionary models. The combination of the highest $m_V$ and $\chi$ with overshooting produces the deepest penetration of the convective zone and the maximum contraction.\par

      
      
      
      
      
      

\begin{table}
  \begin{center}
    \caption[Physical properties of the core at bump]{ Helium core mass, logarithm of central temperature ($\rm Log\ T_{c}$) and pp and CNO nuclear reactions energy for models with different dark photon parameters, $\rm 1.0\, M_{\odot}$, $\rm Z_{\odot}$, and $\alpha_{ov} = 0, 0.025$ at the minimum of the RGB bump.}
    \label{tab:table2}
    \begin{tabular}{c|c|c|c|c|c} 
     \textbf{$\alpha$}& \textbf{$\chi$} & \textbf{$\rm m_{V}\ (eV)$} &\textbf{$\rm M_{He}/M_{\odot}$} &\textbf{$\rm Log\ T_c$}& $\rm Log\ L_{nuc}/L_{\odot}$\\
      
      \hline
      
     0 & $0$ & 0& 0.242 &7.5590 & 1.450\\
     
      0& $2\times 10^{-15}$ &900& 0.240&7.5438 & 1.438\\
      
      0&$7\times 10^{-15}$ &900 & 0.229&7.4142 &1.363 \\
      
     0& $9\times 10^{-15}$ &900 &0.224 & 7.3559 & 1.336\\
     0& $5\times 10^{-15}$ &500 & 0.239 & 7.5297  & 1.429    \\
     0& $5\times 10^{-15}$ &700 & 0.236  & 7.5043 & 1.412\\
     0& $5\times 10^{-15}$ &900 &0.234 & 7.4745 &1.395 \\
      0.025 & 0& 0&0.232 & 7.5398&1.325\\
        0.025 & $2\times 10^{-15}$ &900& 0.230 &7.5240 & 1.313 \\
      
      0.025 &$7\times 10^{-15}$ &900 &0.218  &7.3882 & 1.230 \\
      
      0.025& $9\times 10^{-15}$ &900 &0.214 &7.3284  &1.203 \\
     0.025& $5\times 10^{-15}$ &500 & 0.229 & 7.5091 & 1.302    \\
     0.025& $5\times 10^{-15}$ &700 & 0.226  & 7.4822 &1.283 \\
     0.025& $5\times 10^{-15}$ &900 &0.223 & 7.4508 & 1.265 \\
     
    \end{tabular}
  \end{center}
\end{table}

\begin{table}
  \begin{center}
    \caption{Ratio of mean and central densities, $\rm \rho/\rho_{c}$, bottom of the convective envelope in lagrangian ($\rm M/M_{\odot}$) coordinates  , and maximum value of $\Delta \delta_{2}\nu_{0n}$  for different dark photon models with $\rm 1.0\ M_{\odot}$ and $\rm Z_{\odot}$ at the minimum of the RGB bump.}
    \label{tab:table3}
    \begin{tabular}{c|c|c|c|c|c} 
      \textbf{$\alpha$} & \textbf{$\chi$} & \textbf{$\rm m_{V}\ (eV)$}& \textbf{$\rm \rho/\rho_{c}\, ( \times 10^{8})$} & C. bottom &\textbf{$\rm \delta_{2}\nu_{0n}\ (\mu Hz)$}\\
      
      \hline
      
       0&$ 0$ & $0$ & 1.27 &0.264& 0.3573\\
      
      0& $2\times 10^{-15}$ &900&1.33 &0.262 & 0.3835\\
      
     0& $7\times 10^{-15}$ &900& 1.79 & 0.251&0.7108 \\
      
     0& $9\times 10^{-15}$ &900& 1.98  &0.247 &  0.3882\\
     0& $5\times 10^{-15}$ &500 & 1.38  & 0.261& 0.4256\\
     0& $5\times 10^{-15}$ &700 &  1.48 & 0.259& 0.4982\\
     0& $5\times 10^{-15}$  &900& 1.58  & 0.256 & 0.5747\\
      0.025 &0 & 0&2.35 & 0.258 &0.3920\\
      0.025& $2\times 10^{-15}$ &900&2.47 &0.257& 0.3890\\
      
     0.025& $7\times 10^{-15}$ &900&3.41 & 0.246& 0.3744 \\
      
     0.025& $9\times 10^{-15}$ &900& 3.79  &0.242 & 0.4376 \\
     0.025& $5\times 10^{-15}$ &500 & 2.58 & 0.256& 0.3822 \\
     0.025& $5\times 10^{-15}$ &700 &  2.77 & 0.263& 0.3612 \\
     0.025& $5\times 10^{-15}$  &900& 2.98  & 0.251 &0.3275  \\
    \end{tabular}
  \end{center}
\end{table}
\section{The pulsation models}
In order to understand the effects of dark photons and overshooting in stellar oscillations, we computed pulsation from our equilibrium models. The aim is to find an asteroseismic observable that can be compared to observations.\par
We used the oscillation code GYRE \citep{2018MNRAS.475..879T,2013MNRAS.435.3406T}, capable of computing adiabatic and non-adiabatic oscillations. Solar-like pulsations are correctly modelled with the adiabatic solution, so we calculated only adiabatic modes. We studied $30$ radial orders (with $n=1,30$ and $\ell=0$), evaluating the large separation and the second difference (Equation \ref{eq:delta02}). The $<\Delta\nu>$ is an average of the large separation differences obtained around the frequency of maximum power, $\nu_{\rm max}$, which depends on the mass, radius and temperatures of each model at RGB bump \citep{2017MNRAS.470.2069G}.\par
\begin{figure}
\includegraphics[width=0.50\textwidth]{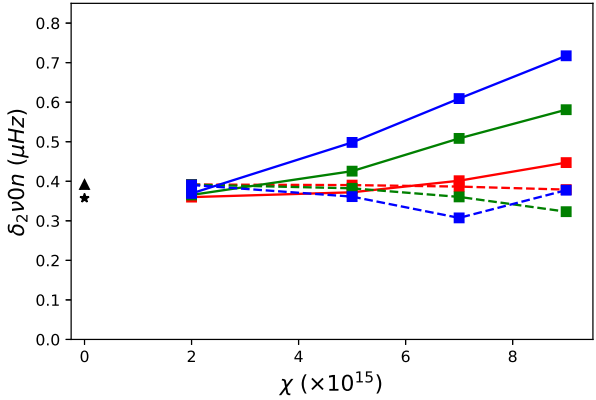}
\caption[$\Delta\delta_{2}\nu_{0n}$]{Variation of $\delta_{2}\nu_{0n}$ vs $\chi$ for $\rm m_{V}=300\ eV$ (red), $\rm m_{V}= 500\ eV$ (green) and $\rm m_{V}= 700\ eV$ (blue). The dashed lines correspond to models with the same parameters and $\alpha_{ov} = 0.025$. The values for a reference model and a pure overshooting model are represented by the black star and the triangle, respectively. In all cases the inital mass and metalicities were $\rm 1\ M_{\odot}$ and
$Z_{\odot}$.}
\label{fig:delta}
\end{figure}

The results for $\delta_{2}\nu_{0n}$ can be seen in Table~\ref{tab:table3} and in Figure~\ref{fig:delta}, where in both cases we show the maximum values of $\delta_{2}\nu_{0n}$ for the $\rm 1 M_{\odot} $ models. In Figure~\ref{fig:delta}, it is observed that, when the combined effects of dark photons and overshooting are considered, the variations of $\delta_{2}\nu_{0n}$ remain in a narrow band. On the contrary, the models where only dark photons are taken into account undergo higher variations, when $\chi$ and $m_{V}$ increase, in agreement with the previously discussed changes in the structure of the core \citep{1994A&A...290..845L}.\par
This behaviour can be explained on the following basis. The dark photons are produced in the RGB degenerate core and $\delta_{2}\nu_{0n}$ is sensitive to the structural changes inside this region. We found that over a certain value, overshooting can increase the extension of the convective envelope notably. That implies that the radiative core size is highly reduced and therefore the structural changes due to dark photons have not an appreciable effect on $\delta_{2}\nu_{0n}$.\par
It is worth to notice that the lack of appreciable deviations of $\delta_{2}\nu_{0n}$ does not exclude the existence of dark photons, as they could yet occur inside stars, but their influence on $\delta_{2}\nu_{0n}$ would be diminished by the envelope overshooting. However, due to their different impacts on $\delta_{2}\nu_{0n}$, dark photons and overshooting could be disentangled, with the aid of other observables. Although in this work, we have focused on the average large separation in order to impose constraints on dark photons, we remark that in conjunction with $\delta_{2}\nu_{0n}$, the average large separation could be used to obtain bounds on overshooting and dark photons simultaneously.\par 

\begin{figure}
\includegraphics[width=0.56\textwidth]{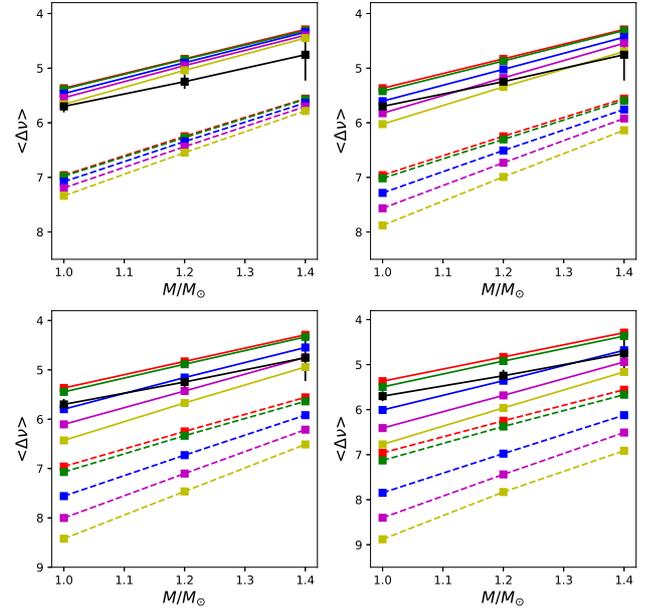}
\caption[The average great separation, $<\Delta \nu>$, vs initial mass $\rm M/M_{\odot}$ for different $\chi$ models]{The average great separation, $<\Delta \nu>$, vs initial mass $\rm M/M_{\odot}$, for models with $m_V$ (dark photon mass) 300 eV (upper left panel), 500 eV (upper right), 700 eV (lower left) and 900 eV (lower right). For each panel there are several models with $\chi=0$ (red), $\chi=2\times 10^{-15}$ (green), $\chi=5\times 10^{-15}$ (blue), $\chi=7\times 10^{-15}$ (magenta), and $\chi=9\times 10^{-15}$ (yellow). The dashed lines represent the introduction of $\alpha = 0.025$ in the models.The theoretical models are represented by squares, and connected by solid lines. The squares with $2\sigma$ error bars over the black line correspond to the values of the observational data. The error bars for $\rm 1.0$ and $\rm 1.2\ M_{\odot}$ are scarcely noticeable.}
\label{fig:averagepanel}
\end{figure}

 Looking into the impact of dark photons in the large separation, we used a well defined sample of RGB stars at the bump. The most recent and completed search for stars in this evolutionary state is that done by \citet{2018ApJ...859..156K}. They used the APOKASC catalogue \citep{0067-0049-233-2-25} together with theoretical predictions of the distribution of $<\Delta\nu>$ and $\nu_{\rm max}$ in order to locate a number of stars at RGB bump. Their conclusions point to a dependence of $<\Delta\nu>$ with the envelope overshooting considered in the modelling. The result of comparing the observations and our models is shown in Figure~\ref{fig:averagepanel}.\par
 
It is noticeable a mismatch between the observations and the predicted values of the average large separations for the  oveshooting-only model with $\alpha_{ov} = 0.025$. Although it could seem in disagreement with the results reported by \citet{2018ApJ...859..156K}, it must be pointed out that in this work the authors cover the range $\rm -0.1 \leq |Fe/H| \leq 0.1$, and they report that the efficiency of overshooting increases with decreasing metallicity. Due to the fact they obtain an average between different metallicities, there are some discrepancies with respect to the results of our work for the overshooting model, which is calculated using the solar metallicity. In addition, some minor offset can arise because Khan et al. used the MESA version 7385, whereas we performed all the calculations with the version 10398. However, we computed models with same parameters definition using MESA v7385 and v10398, and found minimal differences between the values in $<\Delta \nu>$ calculated with both versions. In any case, the offset does not affect the self-consistency of our results and allow us to  derive confident conclusions about the combined effects of dark photons and overshooting.\par 
 
 As it can be observed in Figure~\ref{fig:averagepanel} and in Table~\ref{tab:tableovDnu}, where we show the $<\Delta \nu>$ values for some of the $\rm 1 M_{\odot}$ models, the effect of dark photons is an increase of the large separation with respect to the reference model as $m_V$ is enhanced. This effect is augmented when higher values of $\chi$ are considered, becoming greater than the observational uncertainties of $<\Delta \nu>$ for RGB stars, $\rm 0.2\ \mu Hz$ at $2\sigma$ \citep{2010A&A...517A..22M}. It is remarkable that the reference models without dark photons nor overshoot are not able to reproduce the observations. On the other hand, by means of including dark photons in the models the differences between predictions and observations for the entire mass set can be reduced. In fact, both the envelope overshooting and dark photons increase the large separation prediction.\par

\begin{figure}
\includegraphics[width=0.45\textwidth]{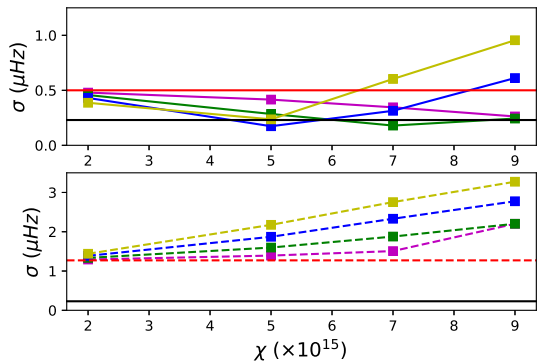}
\caption[Standard deviation for different dark photon models]{Standard deviation between the expected and observed $\rm <\Delta\nu>$ for the different models. The solid lines with squares represent the scenarios without overshooting, whereas the dashed with squares are the models combining dark photons and overshooting. The colors correspond to $\rm 300\ eV$ (magenta), $\rm 500\ eV$ (green), $\rm 700\ eV$ (blue) and $\rm 900\ eV$ (yellow). The horizontal black solid lines correspond to the observational $\rm 2 \sigma$ errors. The red solid one corresponds to the standard deviation of the reference model, without dark photons, whereas the red dashed indicates the standard deviation for the $\alpha_{ov} = 0.025$ model.}
\label{fig:sd}
\end{figure}

In order to estimate the errors of the predicted <$\Delta \nu$> with respect to the observational data, we computed the standard deviation, $\sigma$, for all the models. A plot of $\sigma$ vs $\chi$ and $m_{V}$ , is shown in Figure ~\ref{fig:sd}. The values of $\sigma$  for some models are shown in Table~\ref{tab:tableovsd}. \par

Even considering the combined effect of dark photon and overshooting, some parameters of the dark photon models can be ruled out, like the couple $\rm m_{V}= 900\ eV$, and $\chi = 5\times 10^{-15}$, which correspond to an upper bound. The other excluded models are shown in Table~\ref{tab:tableovexcl}. These models can be discarded  on the basis of the increase of the large separation that they produce. As models like these without overshooting are already out of the 2$\sigma$ error bars of the measured <$\Delta \nu$>, adding even a small fraction of overshooting acts moving the models away. This gives the important consequence that we can constrain the parameter space of dark photon with Asteroseismology even when other ingredients in modelling are uncertain. Moreover, taking into account that the error bars of $\Delta \nu$ for the masses $\rm 1.0\ M_{\odot}$ and $\rm 1.2\ M_{\odot}$ are smaller than for $\rm 1.4\ M_{\odot}$, we suggest the possibility of discarding a broader range of $m_{V}$ and $\chi$ values, those that imply a mismatch with respect to the $\rm 1.0\ M_{\odot}$ observational data beyond $2\sigma$ confidence levels.\par

It is worth to notice that previous astrophysics constraints on dark photons present an essential difficulty, since the evolution of a star depends on many physical processes, some of which are not well-defined: having these constraints based uniquely on energy loss by dark photons is not very robust.  In this sense, the present work is a considerable progress in obtaining robust astrophysical constraints for the dark photon model, mainly because the new class of stellar models used in this study is very solid. It uses a very well-built and tested stellar evolution code, a well-known set of input macro and microphysics, and up-to-date astronomical and asteroseismology data sets.  Furthermore, in the present study the stellar model with dark photon physics includes the feedback of the energy loss by a dark photon in the evolution of the RGB, a process not included in previous astrophysics studies.\par

\begin{table}
  \begin{center}
    \caption[$\nu$ and $\Delta \nu$ for different values of dark photon models ($\chi$ and $m_V$), and overshooting ($\alpha$)]{Values of average large separation, $<\Delta \nu>$, for models with different $\alpha$, $\chi$ and $m_V$. This table corresponds only for some 1${\rm M}_{\odot}$ models, for illustrative purpose. The complete table containing 1.2 and $1.4{\rm M}_{\odot}$ is available in electronic format.}
    \label{tab:tableovDnu}
    \begin{tabular}{c|c|c|c} 
      \textbf{$\alpha$}& \textbf{$\chi$} & $\rm m_{V}\ (eV)$  &\textbf{$\rm <\Delta \nu>\ (\mu Hz)$}\\
      
      \hline
      
      $0$ &0 & 0.00 &  5.36 \\
      $0.025$ &0 & 0.00 &6.96 \\
      0& $2\times 10^{-15}$ &900&5.50 \\
      
     0& $7\times 10^{-15}$ &900&6.41 \\
      
     0& $9\times 10^{-15}$ &900&6.77 \\
     0& $5\times 10^{-15}$ &500 &5.60 \\
     0& $5\times 10^{-15}$ &700 &5.80  \\
     0& $5\times 10^{-15}$  &900&6.01  \\
      0.025& $2\times 10^{-15}$ &900& 7.13 \\
      
     0.025& $7\times 10^{-15}$ &900&8.40  \\
      
     0.025& $9\times 10^{-15}$ &900& 8.88\\
     0.025& $5\times 10^{-15}$ &500 &7.28  \\
     0.025& $5\times 10^{-15}$ &700 &7.56 \\
     0.025& $5\times 10^{-15}$  &900&7.84 \\

     \end{tabular}
  \end{center}
\end{table}

\begin{table}
  \begin{center}
    \caption[Values of standard deviations with respect to the observations for models with different $\alpha$, $\chi$ and $m_V$]{Values of standard deviations with respect to the observations for models with different $\alpha$, $\chi$ and $m_V$. The observational $2\sigma$ limits are $\rm 0.23\ \mu Hz$}
    \label{tab:tableovsd}
    \begin{tabular}{c|c|c|c} 
      \textbf{$\alpha$}& \textbf{$\chi$} & $\rm m_{V}\ (eV)$  &\textbf{$\rm \sigma$ ($\rm \mu Hz$)}\\
      
      \hline
      
      $0$ &0 & 0.00 &0.50 \\
      $0.025$ &0 & 0.00 &1.27\\
      0& $2\times 10^{-15}$ &900&0.39\\
      
     0& $7\times 10^{-15}$ &900& 0.60\\
      
     0& $9\times 10^{-15}$ &900&0.96\\
     0& $5\times 10^{-15}$ &500 &0.28\\
     0& $5\times 10^{-15}$ &700 &0.17\\
     0& $5\times 10^{-15}$  &900&0.24\\
      0.025& $2\times 10^{-15}$ &900& 1.44\\
      
     0.025& $7\times 10^{-15}$ &900&2.75\\
      
     0.025& $9\times 10^{-15}$ &900& 3.27\\
     0.025& $5\times 10^{-15}$ &500 &1.60\\
     0.025& $5\times 10^{-15}$ &700 &1.87\\
     0.025& $5\times 10^{-15}$  &900&2.18\\

     \end{tabular}
  \end{center}
\end{table}

\begin{table}
  \begin{center}
    \caption[Models excluded in any overshooting scenario]{List of dark-photon properties (mass and coupling constant, respectively) discarded whatever the stellar mass and overshooting considered. Models are discarded when do not predict the observed large separation within $2\,\sigma$ error.}
    \label{tab:tableovexcl}
    \begin{tabular}{c|c|c} 
        $\rm m_{V}\ (eV)$ & \textbf{$\chi$}  &\textbf{$\rm \sigma$ ($\rm \mu Hz$)} \\
      
      \hline
      
       900 & $5\times 10^{-15}$ & 0.24 \\
       900 & $7\times 10^{-15}$ & 0.60\\
       900 & $9\times 10^{-15}$ & 0.96\\
       700 & $7\times 10^{-15}$ & 0.31\\
       700 & $9\times 10^{-15}$ & 0.61\\
       500 & $9\times 10^{-15}$ & 0.24\\
     \end{tabular}
  \end{center}
\end{table}    

\section{Dark photon effects on the RGB tip}

In the last step in our work we show the possibility of extending our analysis to another observable of the RGB phase, the RGB-tip luminosity, which corresponds to a local maximum in HR diagram by the He-flash. This observable has been proved to be strongly dependent on the He-core mass, which should be affected by the dark photon cooling, considering the change of the He mass verified previously at the bump for the dark photon models. \par

In Table ~\ref{tab:tablehef} we show the variations of He core mass and luminosity at the RGB tip for two different dark photon models, as well as a couple of models, without overshooting and with $\alpha_{ov}=0.025$. The He core mass increases with $\chi$ and $m_v$, which implies an increase of the luminosity at the RGB tip.\par 

The variations of luminosity we predict, correspond to a variation of magnitude higher than the observational uncertainties for the TRGB of the globular cluster M5. Our theoretical model with $m_{V} = 900$ and $\chi = 9\times 10^{-15}$ gives a $\rm log\ L/L_0$ increase of $\thickapprox 0.1$. This variation correspond to a magnitude variation $\thickapprox 0.3$, which is more than twice the observational uncertainties ($\rm 0.12\ mag$) reported by \citet{2017A&A...606A..33S} and \citet{2013PhRvL.111w1301V}.\par 

As overshooting leaves unaltered the He mass and practically the luminosity at the RGB tip, this may help to constrain the dark photon properties when studied simultaneously with the RGB bump (a work now in progress).\par
 
\section{Conclusions and future work}
In this work we have constrained the parameter space of dark photons using asteroseismic observables. For the first time as far as we know, we have looked into the influence of these particles on the Asteroseismology of stars different to the Sun (placed at the RGB bump), with a focus on the changes in the large frequency separation and the photometric properties of the RGB.\par
We have found that, at the RGB bump, the dark photons modify the internal structure and the depth of the convective borders, as well as produce an increase of density. This affects the second difference parameters of the models and, as we have verified, the average large separation. \par
Using the same observational sample as \citet{2018ApJ...859..156K} and comparing the predicted values of $<\Delta \nu>$ with observational data from the APOKASC catalogue we studied the combined effects of the relatively poor constrained envelope overshooting and dark photons cooling on $<\Delta \nu>$ at the RGB bump, finding that both processes reinforce each other. Even though, some combinations of $\chi$ and $\rm m_{V}$ can be discarded because the models without overshooting are already out of observations. It is so expected that a better calibration of the overshooting at this evolutionary stage would give tighter constraints in the dark-photons parameter space. Regarding the second difference parameter, it decouples the effects of overshooting from those of dark photons. Therefore, combining the second difference with $<\Delta \nu>$ would allow us to put better constrains simultaneously on both quantities: overshooting and dark photons.

Moreover, the RGB tip luminosity (TRGB) is not modified by the different overshooting values. On the contrary different dark photon models affect the helium core mass at TRGB, and consequently the luminosity maximum, producing a magnitude enhancement around $\rm 0.3\ mag$, which is higher than the observational uncertainty for Globular Clusters stars. Hence the TRGB can complement the observational information about the bump in order to look into the possible dark photon cooling and constrain the dark photon models. This will be explore in detail in an upcoming work.\par

The constraint obtained in this work using asteroseismology data is an improvement in relation to the previous astrophysical constraints (dark photon with mass between $\rm 300\ eV$ and $\rm 900\ eV$ and with a kinematic mixing larger that $2-9 \times 10^{-5}$).  This is so, because feedback of the energy loss is included in the evolution of the star, but even more importantly because the high-quality data coming from asteroseismology allows us to have a snapshot of the stellar internal structure today. Unlike previous astrophysical constraints, we can guaranty that the internal structure of the star (which can be strongly affected by energy loss) is consistent (or not) with current asteroseismic data.\par
Therefore, this article is one of the first works that use high-quality stellar physics models and asteroseismology data to put robust constraints on the free parameters of the dark photon model.
A denser grid of models, covering a wider range of parameters (several metallicities, different initial rotation rates, etc), at both the bump and the RGB tip, could be relevant in determining the real uncertainties in the predicted deviations of the large separation (work in preparation).\par


\begin{table}
  \begin{center}
    \caption[Variation of RGB-tip luminosity of $\rm 1 M_{\odot}$ models with different values of $\alpha$, $\chi$, and $m_{V}$]{Variation of RGB-tip luminosity of $\rm 1 M_{\odot}$ models with different values of $\alpha$, $\chi$, and $m_{V}$.}
    \label{tab:tablehef}
    \begin{tabular}{c|c|c|c|c|c} 
      \textbf{$\rm Stellar\ mass$} & $\alpha$ &\textbf{$\rm m_V\ (eV)$}&\textbf{$\chi$} & \textbf{$\rm M_{He}$} &\textbf{$\rm Log\ L/L_{\odot}$}\\
      
      \hline
      
      1.0 &0.0 &0 &0& 0.466&3.391 \\
      1.0 &0.025 &0 &0 &0.466 &3.394 \\
      1.0 &0.025 &700 &$5\times 10^{-15}$ &0.474 & 3.435\\
      1.0 &0.0 & 900& $9\times 10^{-15}$&0.491 &3.526\\
      
      \end{tabular}
  \end{center}
\end{table}


\section*{Acknowledgements}

We are grateful to Dr. Saniya Khan and Dr. Andr\'es Moya for useful explanations and advice about the Asteroseismology of RGB stars, as well as to Diane Haun for improving the quality of the text. We are also thankful to the anonymous referee for many valuable suggestions that helped to improve the manuscript. JCS acknowledges funding support from Spanish public funds for research under projects ESP2017-87676-C5-2-R, and from project RYC-2012-09913 under the 'Ram\'on y Cajal' program of the Spanish Ministry of Science and Education. AGH acknowledges funding support from Spanish public funds for research under projects ESP2017-87676-C5-2-R and ESP2015-65712-C5-5-R of the Spanish Ministry of Science and Education.




\bibliographystyle{mnras}
\bibliography{main} 




\appendix




\bsp	
\label{lastpage}
\end{document}